\documentclass[12pt]{iopart}

\usepackage{graphicx}
\usepackage[figuresright]{rotating}

\begin{document}

% declarations for front matter
\title{Higher Flow Harmonics in Heavy Ion Collisions from STAR }

\author{Paul Sorensen (for the STAR Collaboration)}
\address{Brookhaven National Laboratory, 
    Upton, New York 11973-5000, USA}%
\ead{prsorensen@bnl.gov}
       
% If you use the option headings,
% the title is also used as the running title,
% and the authors are also used as the running authors.
% You can change that by using \runtitle and \runauthor.

%\runtitle{$v_n$ and $v_n$ fluctuations}
%\runauthor{P. Sorensen}

\begin{abstract}
We report STAR measurements relating to higher flow harmonics including the centrality dependence of two- and four-particle cumulants for harmonics 1 to 6. Two-particle correlation functions vs. $\Delta\eta$ and $\Delta\phi$ are presented for $p_T$ and number correlations. We find the power spectra (Fourier Transforms of the correlation functions) for central collisions drop quickly for higher harmonics. The $\Delta\eta$ dependence of $v_3\{2\}^2$ and the $p_T$ and centrality dependence of $v_2$ and $v_3$ are studied. Trends are conistent with expectations from models including hot-spots in the initial energy density and an expansion phase. We also present $v_3$ and $v_2\{2\}^2-v_2\{4\}^2$ vs. $\sqrt{s_{_{NN}}}$.  \end{abstract}

\section{Introduction}
Studying the conversion of coordinate space anisotropies into momentum space anisotropies has provided important insight into the characteristics of the matter created in heavy-ion collisions~\cite{reviews}. Elliptic flow ($v_2=\langle\cos2(\phi-\Psi_{RP})\rangle$) has been studied for decades to probe the conversion of the elliptic shape of the initial overlap zone into momentum space~\cite{v2papers}. In 2007, Mishra et. al. proposed the analysis of $v_n^2$ for all values of $n$ as an analogous measurement to the Power Spectrum extracted from the Cosmic Microwave Background Radiation~\cite{Mishra:2007tw}. They argued that density inhomogeneities in the initial state would lead to non-zero $v_n^2$ values for higher harmonics including $v_3$. It was subsequently pointed out that information on $v_n^2$ was to a large extent contained within already existing two-particle correlations data~\cite{Sorensen:2008dm}, and that $v_n$ and $v_n$ fluctuations would provide a natural explanation for the novel features seen in those correlations, such as the ridge like~\cite{ridgedata} and mach-cone like~\cite{awayside} structures. That the ridge could be related to flux-tube like structures in the initial state was already argued by Voloshin in 2003~\cite{radflow}. This was bolstered by hydrodynamic calculations carried out within the NEXSPHERIO model which showed that fluctuations in the initial conditions lead to a near-side ridge correlation and a mach-cone like structure on the away-side~\cite{Takahashi:2009na}. In 2010, Alver and Roland used participant eccentricity ($\varepsilon_{n,\mathrm{part}}$) generalized to arbitrary values of $n$ as introduced in Ref.~\cite{broniowski} and showed that within the AMPT model, the final momentum space anisotropy for $v_3$ was proportional to the initial $\varepsilon_{3,\mathrm{part}}$~\cite{AR}. This explained the previous observation that the AMPT model produced correlations similar to those seen in the data (albeit with smaller amplitudes)~\cite{Ma:2006fm}. Although Mishra et. al. proposed measuring $v_n$ vs. $n$ to observe superhorizon fluctuations, it was subsequently noted that higher harmonics of $v_n$ could be washed out by viscous effects and that the shape of $v_n$ vs. $n$ would provide a tool for studying these effects~\cite{sound}. The effect of viscosity may also make it difficult to observe the effect of the acoustic horizon at higher harmonics.

In this talk, we present measurements from the STAR collaboration~\cite{STAR} relating to two-particle correlations and higher harmonics of $v_n$.  Previous interpretations of two-particle correlations have focused on jet phenomenology~\cite{minijet} and multi-parameter fits adapted from p+p data have been used to succesfully describe heavy-ion data without recourse to higher harmonics. Here we compare correlations and their Fourier Transforms $v_n^2\{2\}$ to expectations from the conversion of density inhomogeneities in the initial conditions into momentum space anisotropies.

\section{Results}

\begin{figure}[htb]
\begin{center}
\resizebox{0.40\textwidth}{!}{\includegraphics{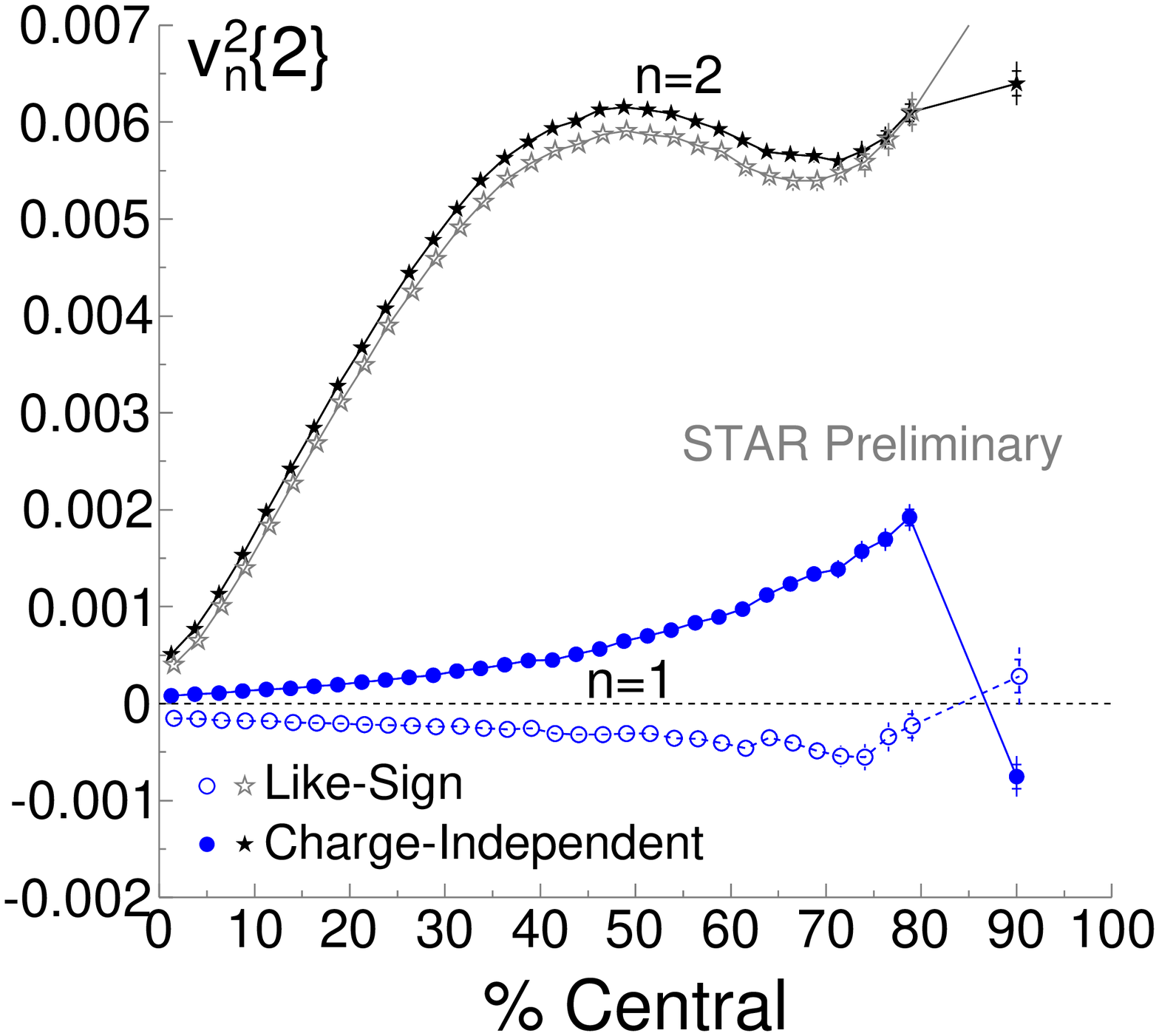}}
\resizebox{0.40\textwidth}{!}{\includegraphics{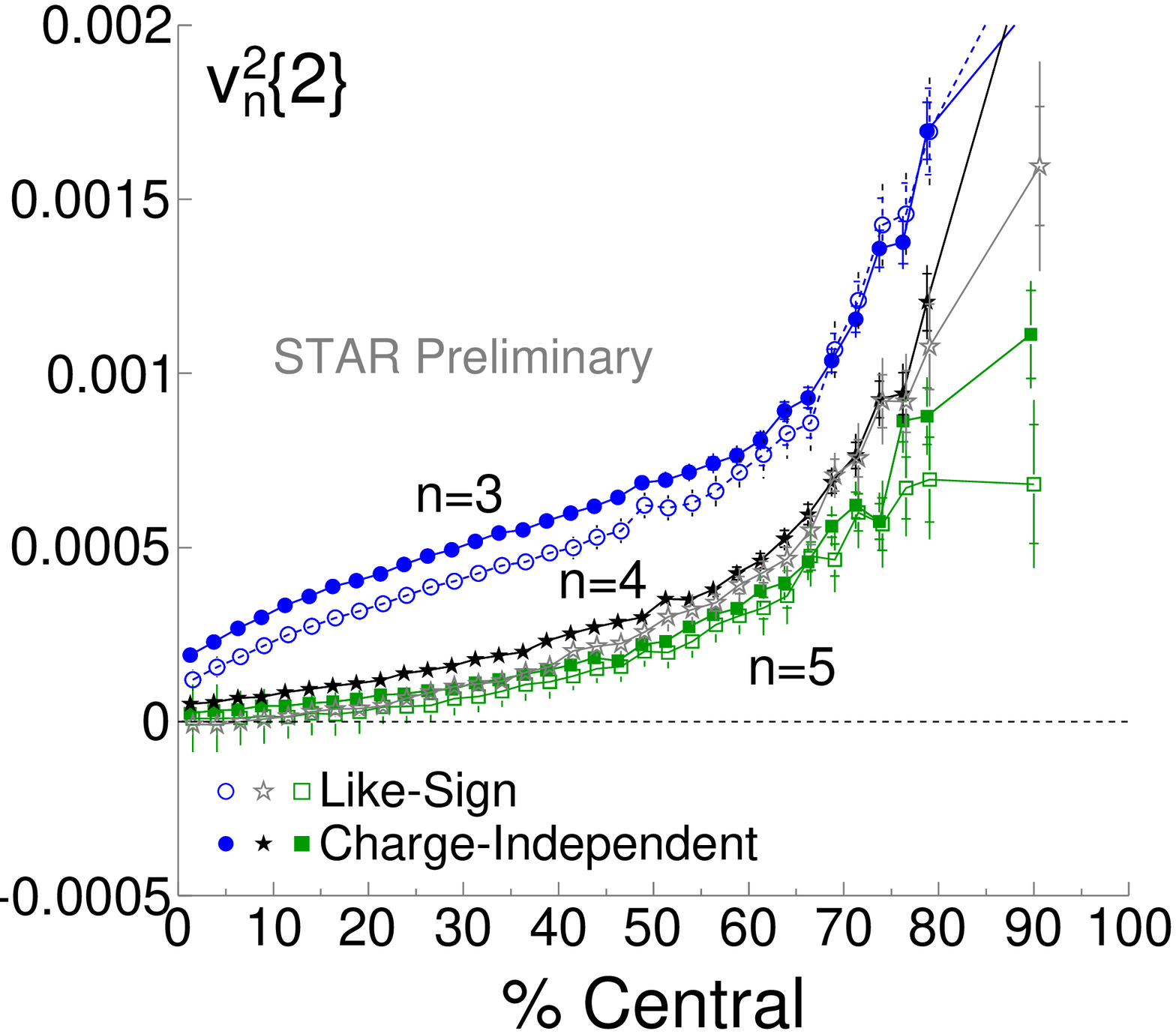}}
\end{center}
\begin{center}
 \resizebox{0.40\textwidth}{!}{\includegraphics{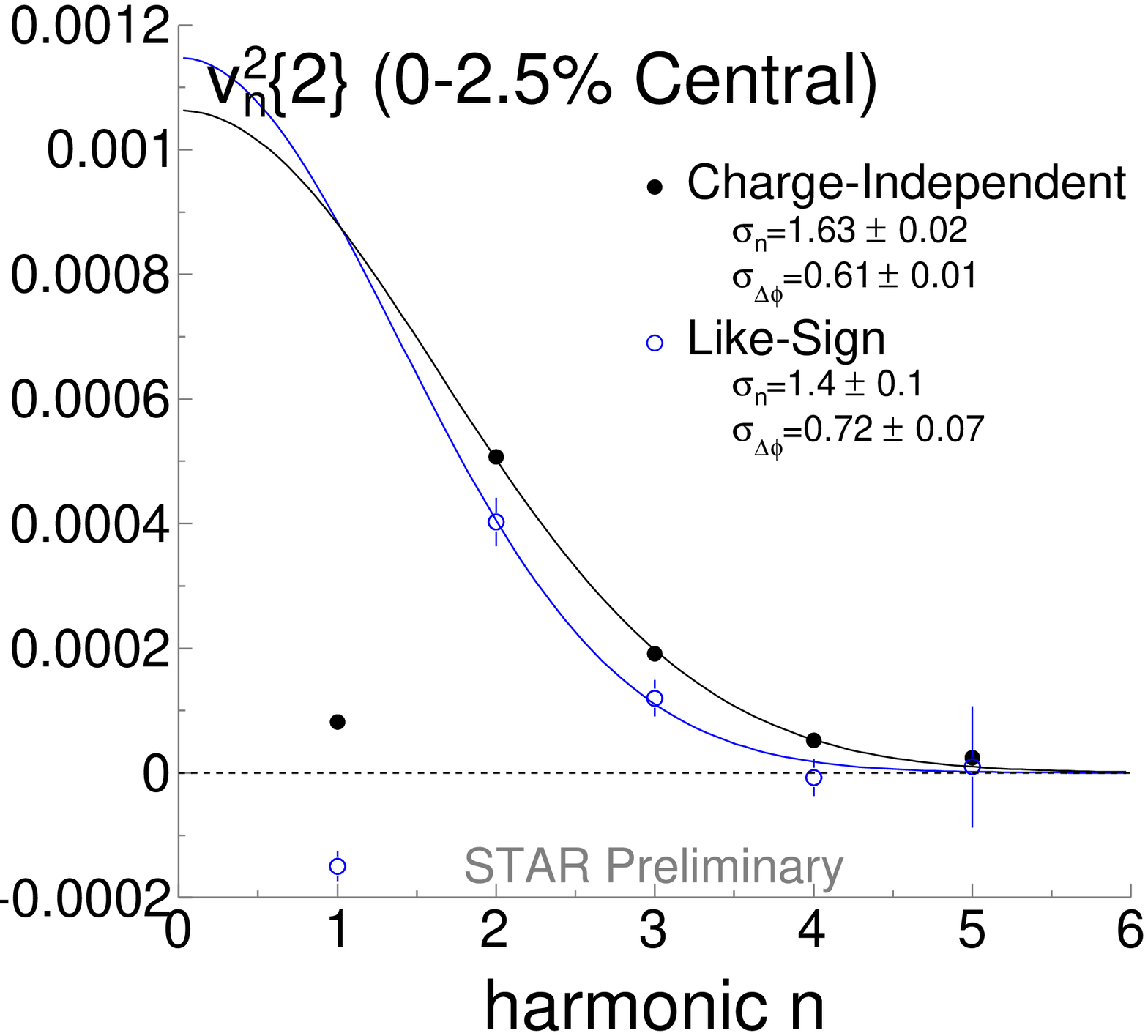}}
\resizebox{0.40\textwidth}{!}{\includegraphics{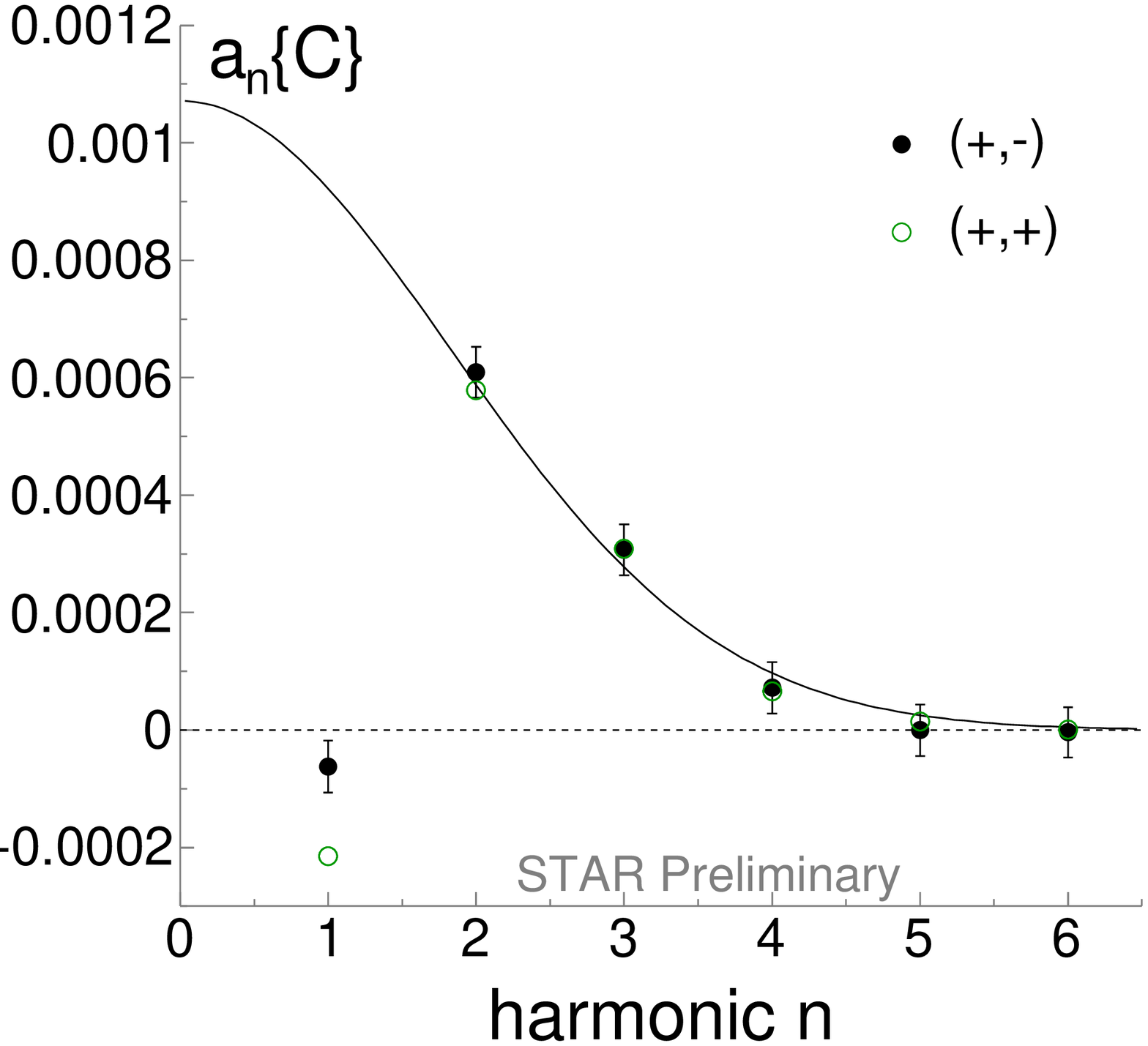}}
\end{center}
\vspace{-0.50cm}
\caption[]{ Top panels: $v_n\{2\}^2$ for 200 GeV Au+Au collisions in 2.5\% centrality intervals. Bottom left panel: $v_n\{2\}^2$ versus $n$ for the 0-2.5\% most central data. Bottom right panel: $a_n$ vs. $n$ from 0-5\% most central $p_T$ correlations with $0.7<\Delta\eta<2.0$. }
\label{f1}
\end{figure}

In Fig.~\ref{f1} (top) we show $v_n\{2\}^{2}$ for $n=1,2,3,4,$ and 5 for 200 GeV Au+Au collisions in 2.5\% centrality intervals using the Q-cumulant method~\cite{qcum} with unit weights. Narrower centrality intervals remove the dependence on weights and bin widths. Results are shown for like-sign (LS) and charge-independent (CI) pairs evaluated with $0.15<p_T<2.0$ and $|\eta|<1$. For $n=1$,  the LS results are negative while the CI results are positive. Positive $v_1\{2\}^2$ indicates particles tend to come out in the same hemisphere while negative indicates particles tend to be more back-to-back. Although $\langle v_1\rangle$ is expected to be symmetric around mid-rapidity and zero at mid-rapidity, fluctuations can lead to deviations from zero from event-to-event and momentum conservation can lead to negative values of $v_1\{2\}^2$. From very peripheral to 70\% central collisions, the $n=2$ term is dropping as expected for non-flow correlations. Beyond 70\% centrality, the effects of elliptic flow overcome the non-flow contributions. The shape of $v_2\{2\}^2$ for 0\%--70\% central is dominated by the elliptic shape of the initial overlap geometry, first rising as multiplicity increases, then decreasing as the overlap region becomes more symmetric. For $n=3$, $v_3\{2\}^2$ in peripheral collisions shows the usual decrease with increasing centrality expected for non-flow correlations which are diluted as $1/N$ as the system volume increases (where $N$ is the number of superimposed independent correlation sources). But for collisions more central than 70-100\%,  $v_3\{2\}^2$ deviates from the typical $1/N$ behavior and exhibits a centrality dependence that echoes  that observed for $n=2$. This is consistent with the almond shape of the initial overlap geometry influencing the $n=3$ term as well~\cite{rise,ty}. For higher n, the deviations from $1/N$ become more difficult to discern. We've measured the four-particle cumulant $v_n\{4\}^{4}$ as well (not shown). Our results for $v_n\{4\}^{4}$ are consistent with zero for all but the $n=2$ term. For the 0-2.5\% most central bin, $v_2\{4\}^{4}$ is also consistent with zero. These results are consistent with expectations for $v_n\propto\varepsilon_{n,\mathrm{part}}$ and a Gaussian approximation for $\varepsilon_x$ and $\varepsilon_y$, where $\varepsilon_{n,\mathrm{part}}^2= \varepsilon_x^2+\varepsilon_y^2$~\cite{gmod}.
% We do not see evidence for the non-gaussian behavior of $\varepsilon_{x,y}$ expected %from Glauber models as shown by the ALICE collaboration at this conference~\cite{}.

The bottom panels of Fig.~\ref{f1} show $v_n\{2\}^2$ (left) and $a_n$ (right) versus $n$ for 0-2.5\% and 0-5\% central collisions respectively. The $v_n\{2\}^2$ results are taken from the Q-cumulant analysis which integrates over all pairs of particles within $|\eta|<1$ and $0.15 < p_T < 2.0$ GeV/$c$. The $a_n$ results are derived from a Fourier Transform of the 2-D $\Delta\eta$ vs. $\Delta\phi$ correlation function $C=\frac{\langle\sum_{i=1}^{n_1}\sum_{j=1\neq i}^{n_2} p_{T,i}p_{T,j} \rangle}{\overline{n}_{1}\, \overline{n}_{2}} - \overline{p}_{T,1}\, \overline{p}_{T,2}$ defined to reflect momentum currents~\cite{Gavin:2006xd,Sharma:2008qr,Sharma:2009zt}. The Fourier Transform is taken for the region $0.7<\Delta\eta<2.0$ thus excluding short range non-flow correlations such as HBT and p+p like jet fragments. The two spectra show the same behavior with the $n=1$ components suppressed and the remainder of the spectra falling off quickly. The drop with $n$ is well described by a Gaussian. The Gaussian width for $v_n\{2\}^2$ vs. $n$ from all pairs independent of charge is $\sigma_{n}=1.63\pm 0.02$ as expected from the observed $\Delta\phi$ width of the near-side peak in the two-particle correlations~\cite{minijet}. If $a_n$ and $v_n\{2\}^2$ are sensitive to the initial eccentricity, $a_1$ and $v_1\{2\}^2$ are suppressed because $\varepsilon_{1,\mathrm{part}}^2\approx 0$ (due to the constraint that the eccentricity is defined relative to the center-of-mass of the participating nucleons~\cite{rise}). Higher harmonics are expected to be small due to viscous effects. The shape of $a_n$ and $v_n\{2\}^2$ are thought to be sensitive probes of viscous effects in heavy ion collisions.

\vspace{-0.0cm}
\begin{figure}[htb]
\begin{center}
\resizebox{0.40\textwidth}{!}{\includegraphics{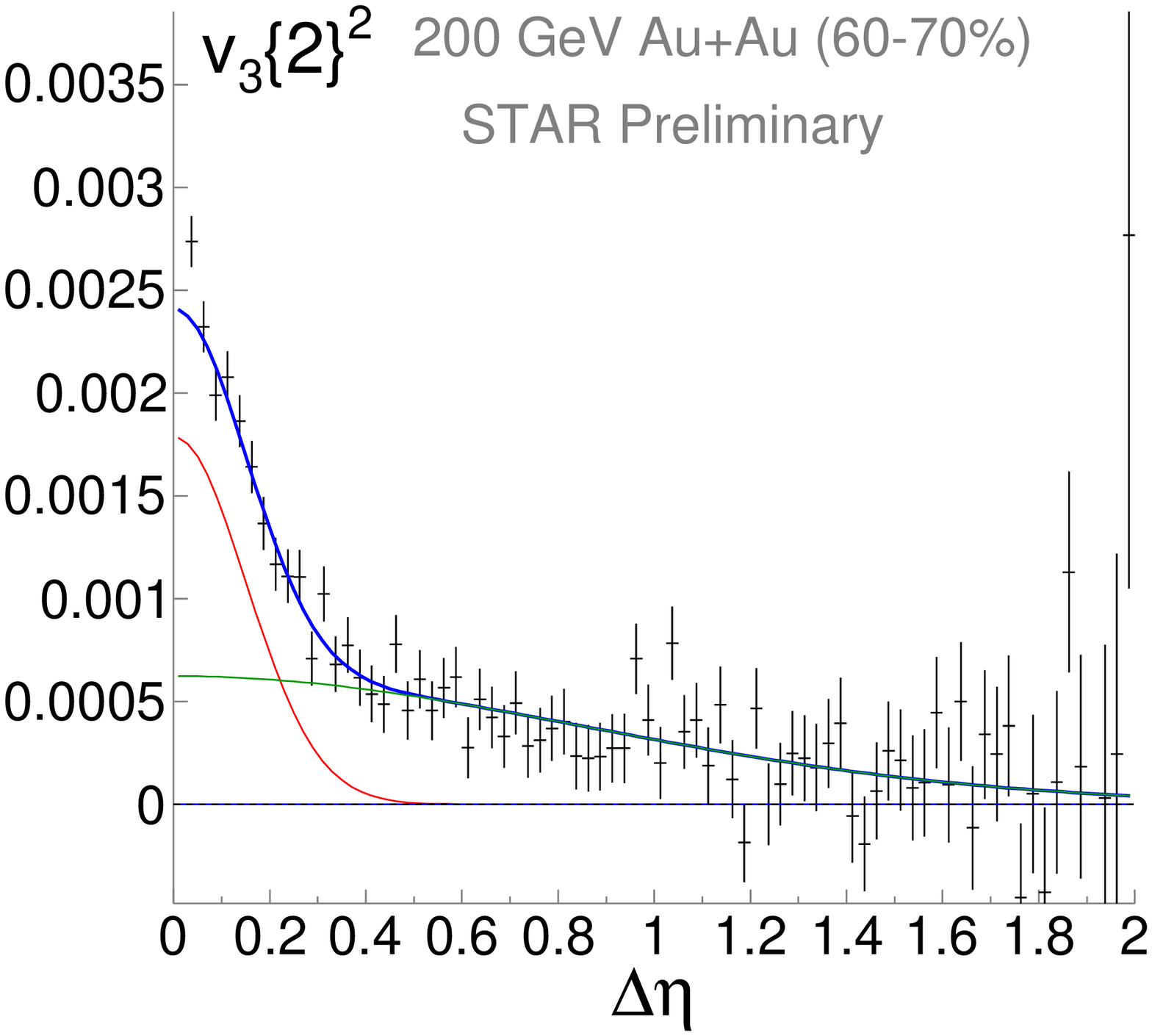}}
\resizebox{0.40\textwidth}{!}{\includegraphics{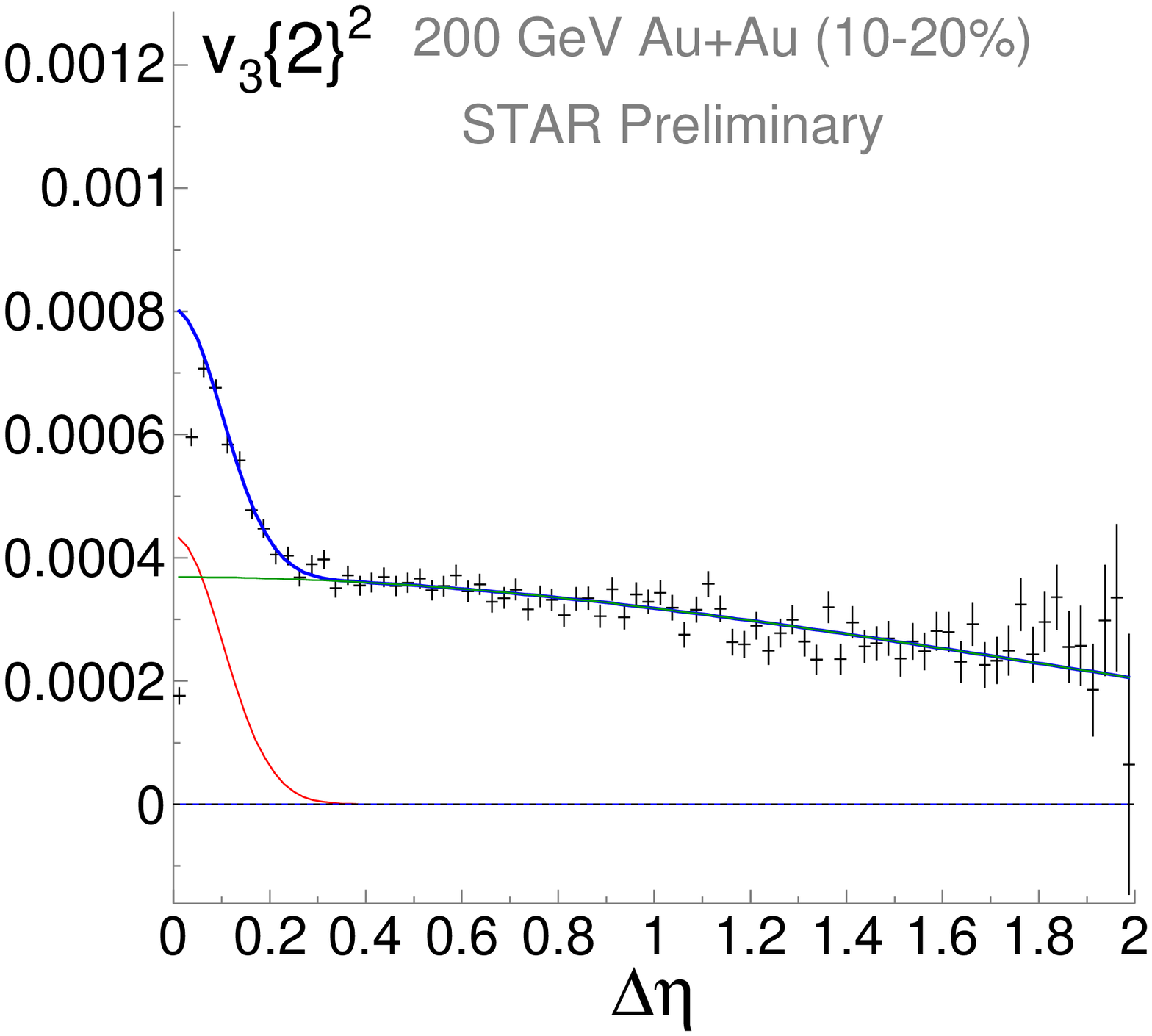}}
\end{center}
\begin{center}
\resizebox{0.40\textwidth}{!}{\includegraphics{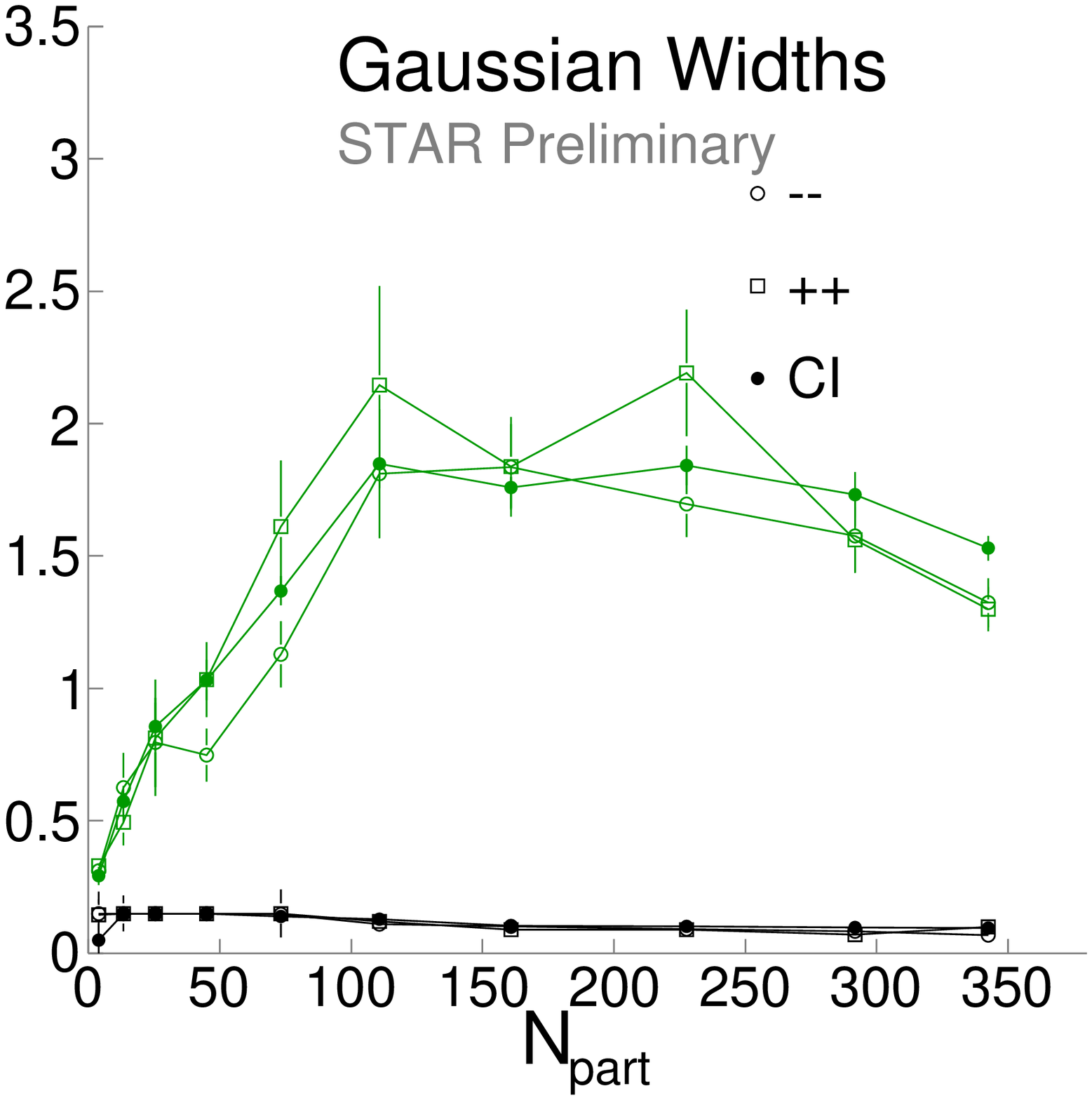}}
\resizebox{0.40\textwidth}{!}{\includegraphics{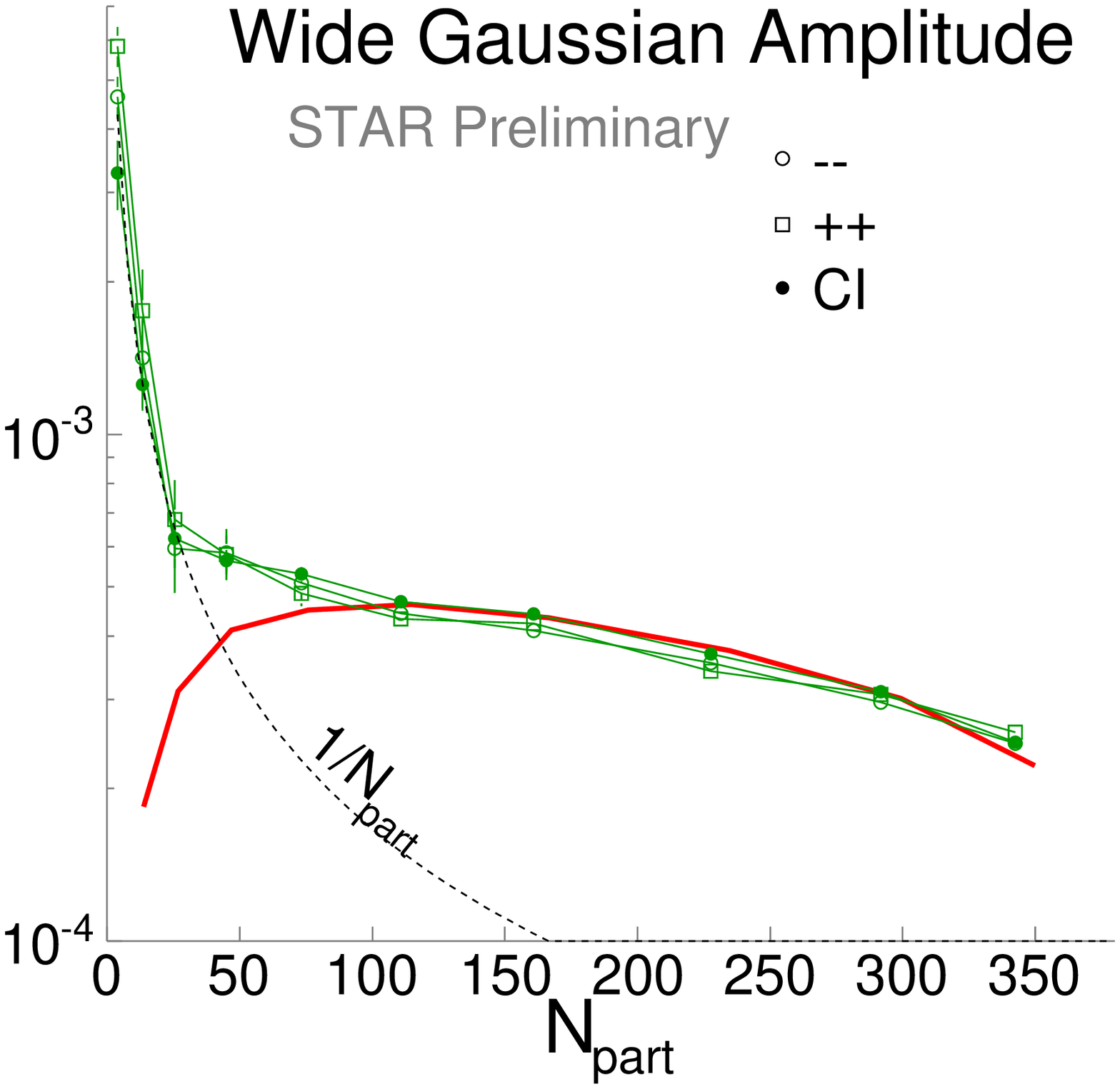}}
\end{center}
\vspace{-0.50cm}
\caption[]{ Top panels: $v_3\{2\}^2$ vs. $\Delta\eta$ for all charged hadrons within two centrality intervals in 200 GeV Au+Au collisions. Data are fit with a narrow and a wide Gaussian. Bottom left panel: The width of the two Gaussians used to fit $v_{3}\{2\}^2(\Delta\eta)$ vs. centrality. Bottom right panel: the amplitude of the wide Gaussian compared to a $1/N_{\mathrm{part}}$ dependence and a $N_{\mathrm{part}}\times\varepsilon_{3,\mathrm{part}}^2$ dependence.}
\label{f2}
\end{figure}

Many sources of correlations besides those related to flow can contribute to two-particle correlations and $v_n\{2\}^2$. More differential analyses can help to reveal the mechanism responsible for the correlations; particularly two-particle correlations vs. $\Delta\eta$ and $\Delta\phi$ and $p_T-p_T$ or $y_T-y_T$ corelations~\cite{Adams:2004pa}. To look at the higher harmonics more differentially we also study $v_3\{2\}^2$ vs. the pseudo-rapidity separation between the two particles  $\Delta\eta$. The top panels of Fig.~\ref{f2} show examples of $v_3\{2\}^2$ vs. $\Delta\eta$ from 200 GeV Au+Au collisions for two centrality intervals, 60-70\% central (top left) and 10-20\% central (top right). $v_3\{2\}^2$ vs. $\Delta\eta$ can be well described by a wide and a narrow Gaussian peak. The narrow Gaussian is consistent with expectations from Bose-Einstein correlations. By mid-central collisions, $v_3\{2\}^2$ is dominated by the wide Gaussian peak. 

In the bottom panels of Fig.~\ref{f2}, we show the width (left) and amplitude (right) of the wide Gaussian for positive, negative, or charge independent particle pairs. The width of the narrow Gaussian decreases with centrality but the width of the wide Gaussian first increases then slightly decreases with centrality. The wide Gaussian reaches a width of $\sigma_{\Delta\eta}=2$ near $N_{\mathrm{part}}=120$ where $N_{\mathrm{part}}$ is estimated using a Glauber Monte Carlo model~\cite{glauber} with a Woods-Saxon distribution for the nuclear density distribution and a 42 mb nucleon-nucleon cross-section. 
%For that width and the precision of the data, the shape of $v_3\{2\}^2$ is
%indistinguishable from a straight line within the STAR acceptance. 
In the bottom right panel, we show the amplitude of the wide Gaussian. The amplitude initially drops as $1/N_{\mathrm{part}}$ with increasing centrality as expected for an independent superposition of non-flow sources. By $N_{\mathrm{part}}=50$, the amplitude deviates sharply from the $1/N_{\mathrm{part}}$ trend. We compare the centrality dependence of the amplitude to the shape of $N_{\mathrm{part}}\times\varepsilon_{3,\mathrm{part}}^2$. We observe that the deviation from the $1/N_{\mathrm{part}}$ trend is well described by the ansatz that $v_{3}\{2\}^2(\Delta\eta=0) \propto N_{\mathrm{part}}\times\varepsilon_{3,\mathrm{part}}^2$ where $\varepsilon_{3,\mathrm{part}}^2$ is calculated from our Glauber Monte Carlo as in Ref~\cite{AR}. The $\Delta\eta$ dependence, however, has not been described in detail by any model we know of. If $v_3\{2\}^2$ is related to the initial eccentricity fluctuations, then the reduction of $v_3\{2\}^2$ at large $\Delta\eta$ would presumably require some decoherence of the initial state fluctuations at large rapidity separations~\cite{Dusling:2009ni}. Recent work has found such a decoherence effect with a hadron and parton cascade model~\cite{Petersen:2011fp}.

\vspace{-0.0cm}
\begin{figure}[htb]
\begin{center}
\resizebox{0.40\textwidth}{!}{\includegraphics{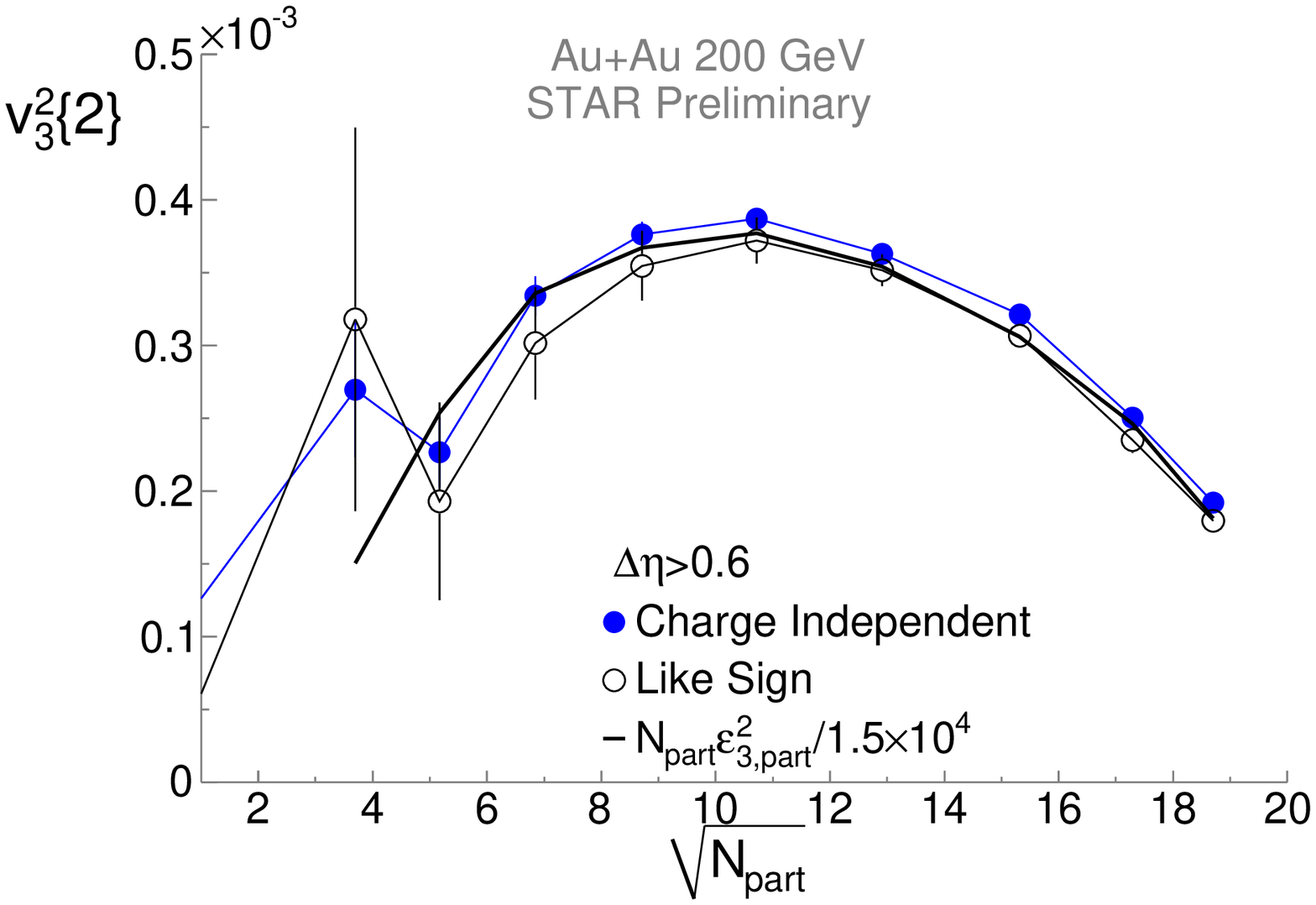}}
\resizebox{0.40\textwidth}{!}{\includegraphics{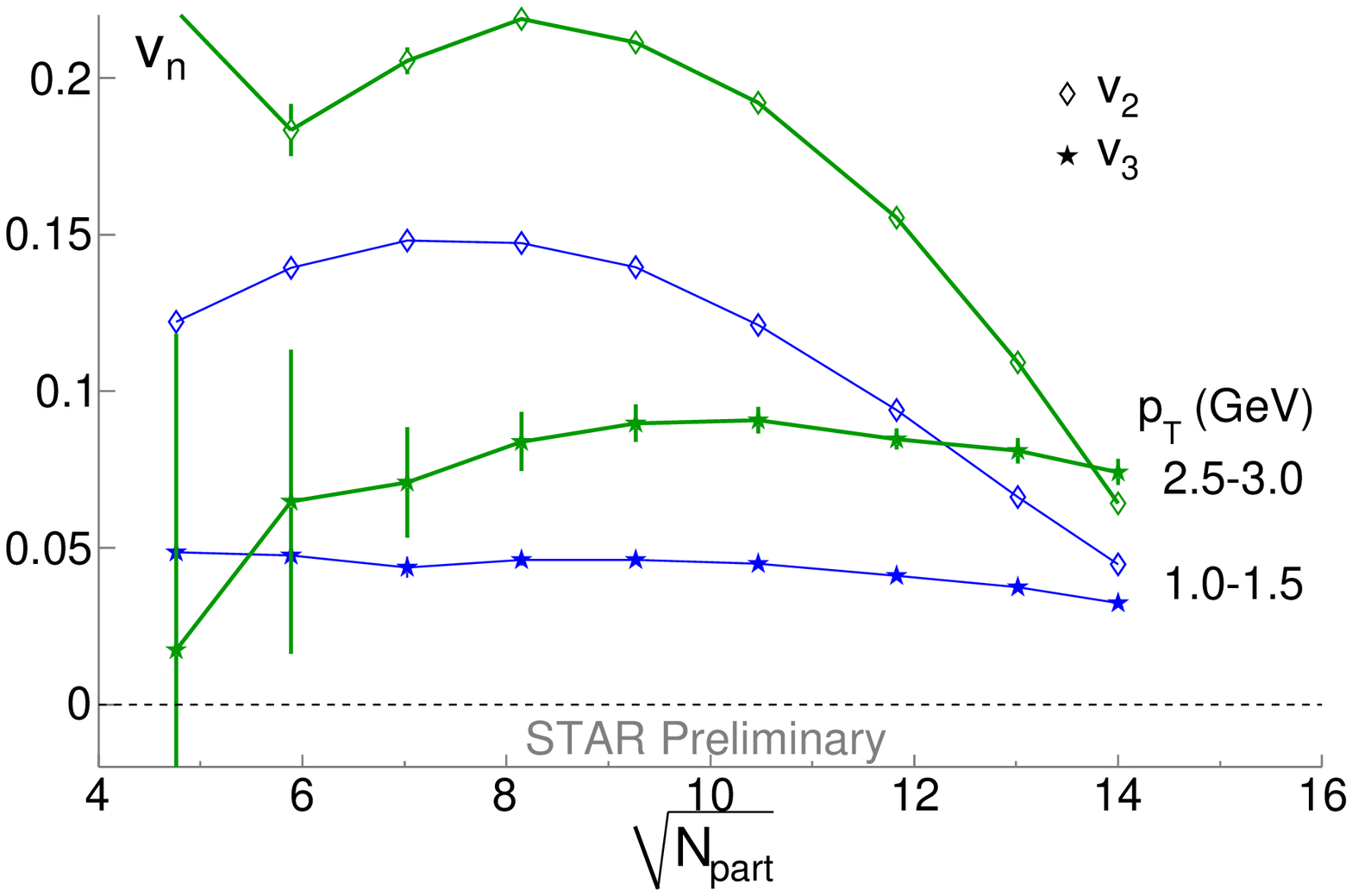}}
\end{center}
\begin{center}
\resizebox{0.40\textwidth}{!}{\includegraphics{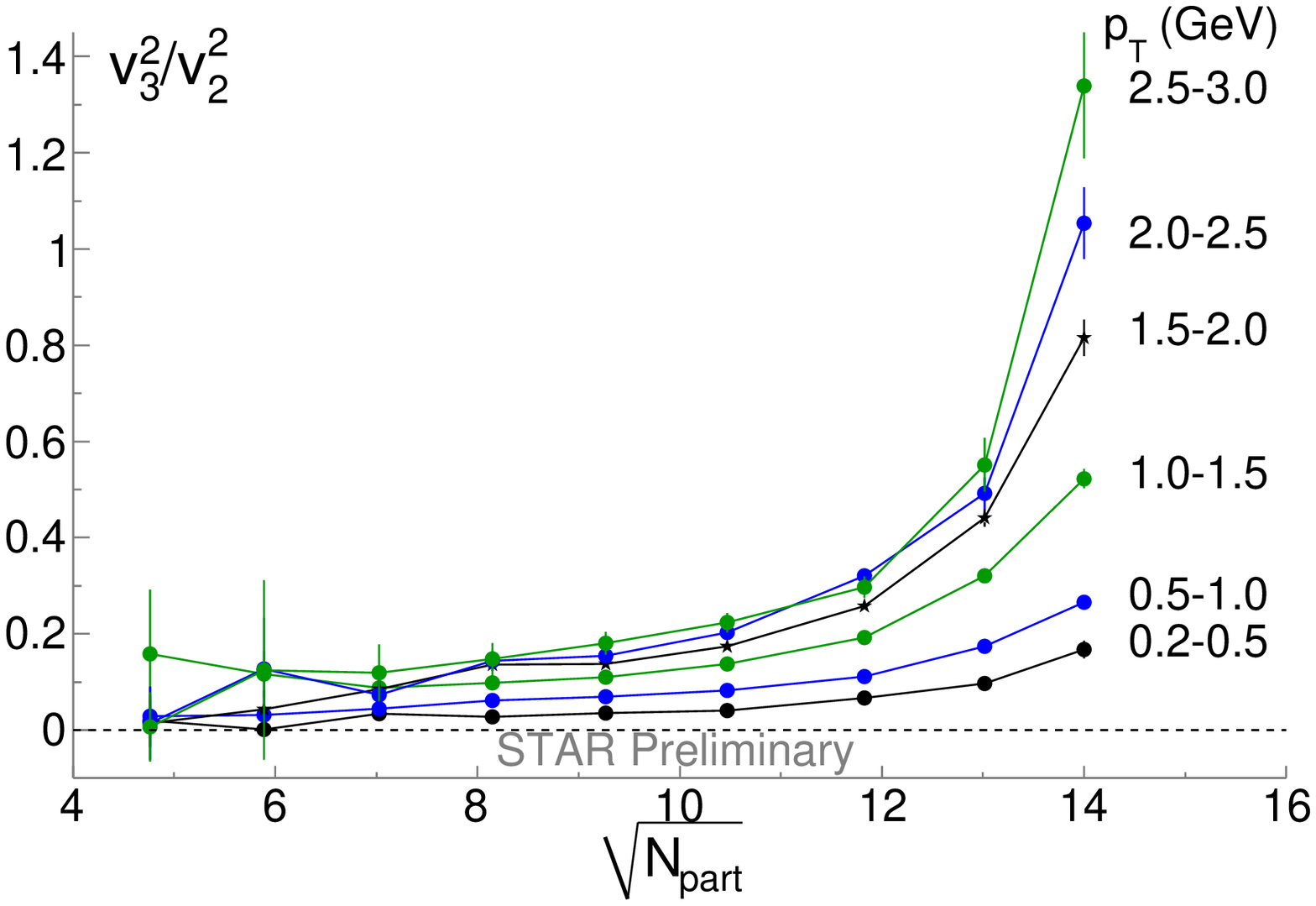}}
\resizebox{0.40\textwidth}{!}{\includegraphics{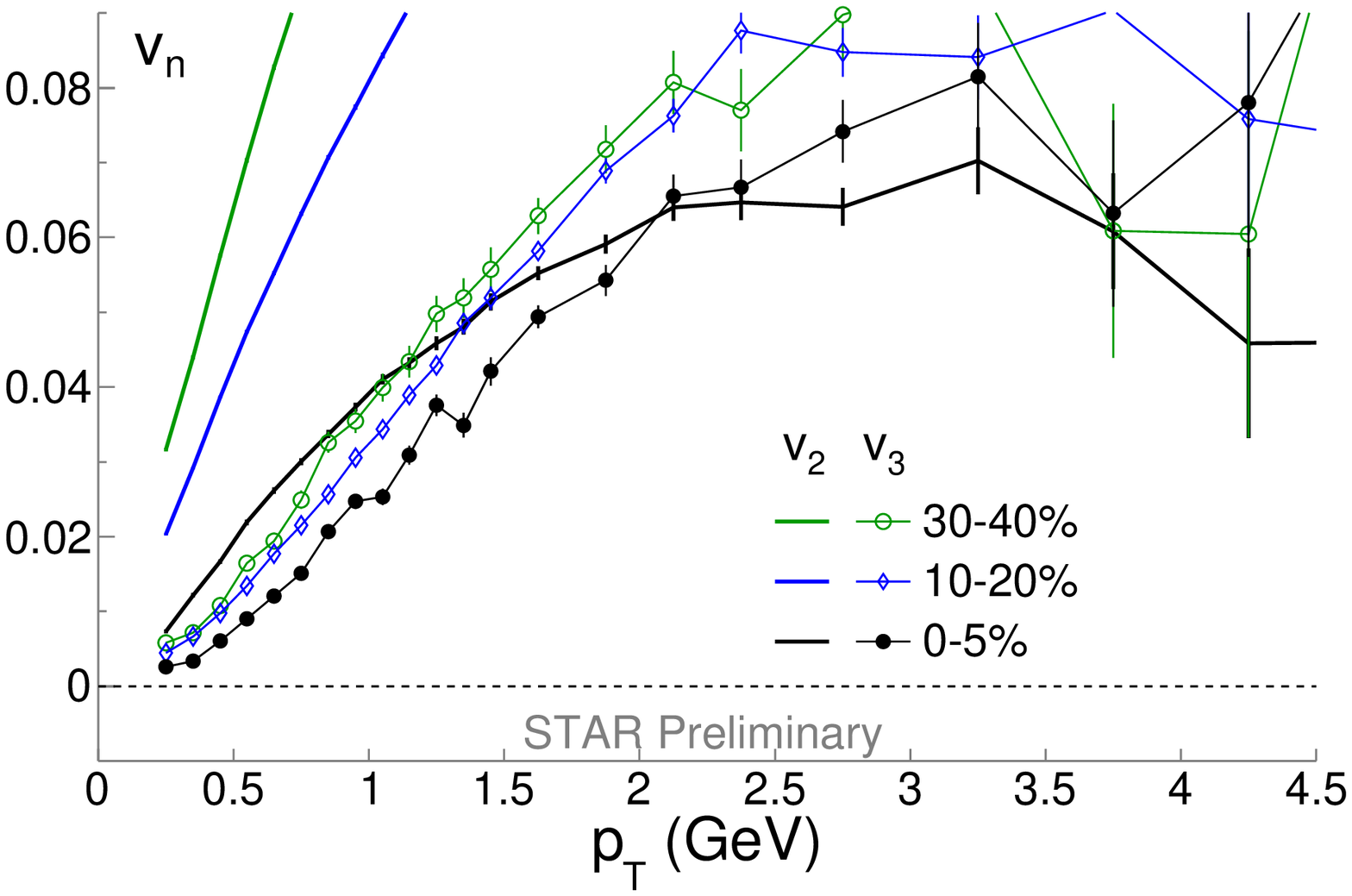}}
\end{center}
\vspace{-0.50cm}
\caption[]{ Top left: the centrality dependence of $v_3\{2\}^2$ for $\Delta\eta>0.6$ as a function of $\sqrt{N_{\mathrm{part}}}$. Top right: $v_2\{2\}$ and $v_3\{2\}$ vs. centrality for two different $p_T$ intervals ($v_n\{2\}$ analyzed using Q-cumulants with Q vectors reconstructed at $\eta>0.5$ and $\eta<-0.5$). Bottom left: $v_3^2/v_2^2$ vs. centrality for different $p_T$ ranges. $v_2\{2\}$ and $v_3\{2\}$ vs. $p_T$ from three different centrality intervals.}
\label{f3}
\end{figure}

In Fig.~\ref{f3} (top left) we compare $v_3\{2\}^2$ for $\Delta\eta>0.6$ to $N_{\mathrm{part}}\times\varepsilon_{3,\mathrm{part}}^2$. The $\Delta\eta>0.6$ cut excludes the narrow Gaussian shown in Fig.~\ref{f2} but introduces a dependence on the Gaussian width. The like-sign, and charge-independent results are compared to $N_{\mathrm{part}}\times\varepsilon_{3,\mathrm{part}}^2$ and found to agree, suggesting $v_3\{2\}^2$ may reflect the initial eccentricity. At top right we plot $v_2\{2\}$ and $v_3\{2\}$ vs. centrality for two $p_T$ intervals (see~\cite{liyi}). These results are measured using the Q-cumulant method with Q vectors reconstructed from different sides of the detector ($\eta<-0.5$ and $\eta>0.5$). This builds in a $\Delta\eta$ gap of at least $\Delta\eta>1$ thus eliminating non-flow from HBT correlations and p+p like jet fragments but nonflow from highly modified jets could persist. $v_2$ first rises as the system becomes more dense, then drops as the overlap region becomes more symmetric. $v_3$ shows a similar but weaker centrality dependence. In the bottom right panel, we show the ratio of $v_3\{2\}^2/v_2\{2\}^2$ from the Q-cumulant analysis with $\eta$ gaps vs. centrality for six $p_T$ intervals. The ratio is smallest in peripheral collisions but then increases with centrality, rising abruptly in the central bins. The ratio also increases with $p_T$ with the most central results in the highest $p_T$ intervals reaching a value greater than unity. This is also the $p_T$ range where 0-1\% central two-particle correlations vs. $\Delta\eta$ and $\Delta\phi$ show clear evidence of higher flow harmonics before background subtraction~\cite{chanaka}.
 Similar behavior was noted before in the ratio of $v_n\{2\}^2/\varepsilon_{n,\mathrm{part}}^2$~\cite{Sorensen:2011xw}, based on STAR correlations data~\cite{Aggarwal:2010rf}. The rise of $v_3$ above $v_2$ was predicted in several models but the origin of the effect has not yet been fully explained. In the bottom right panel we show $v_2$ and $v_3$ vs. $p_T$ for three different centrality intervals. $v_3$ increases with $p_T$ and shows a weaker centrality dependence than $v_2$.

\vspace{-0.0cm}
\begin{figure}[htb]
\begin{center}
\resizebox{0.40\textwidth}{!}{\includegraphics{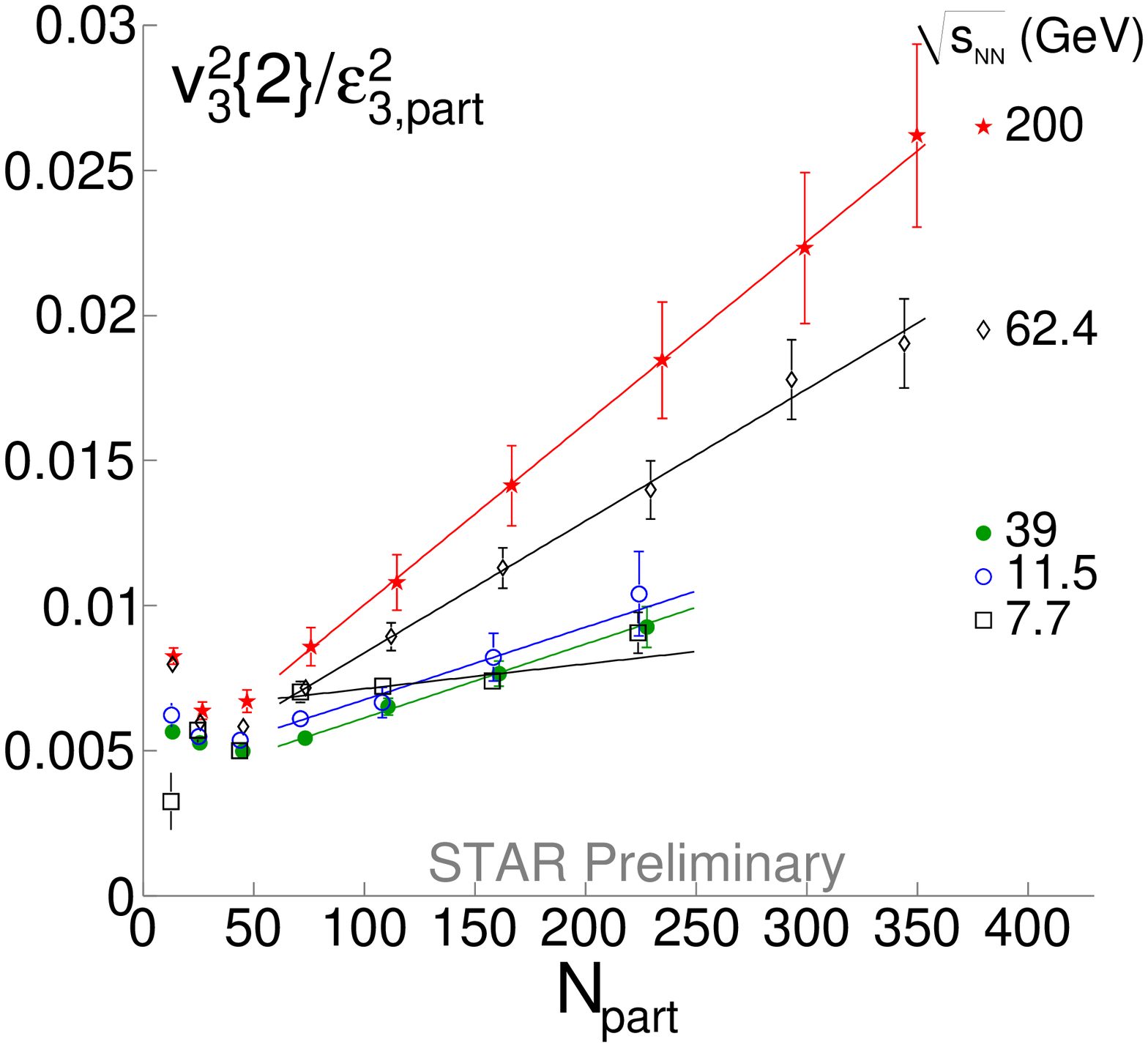}}
\resizebox{0.40\textwidth}{!}{\includegraphics{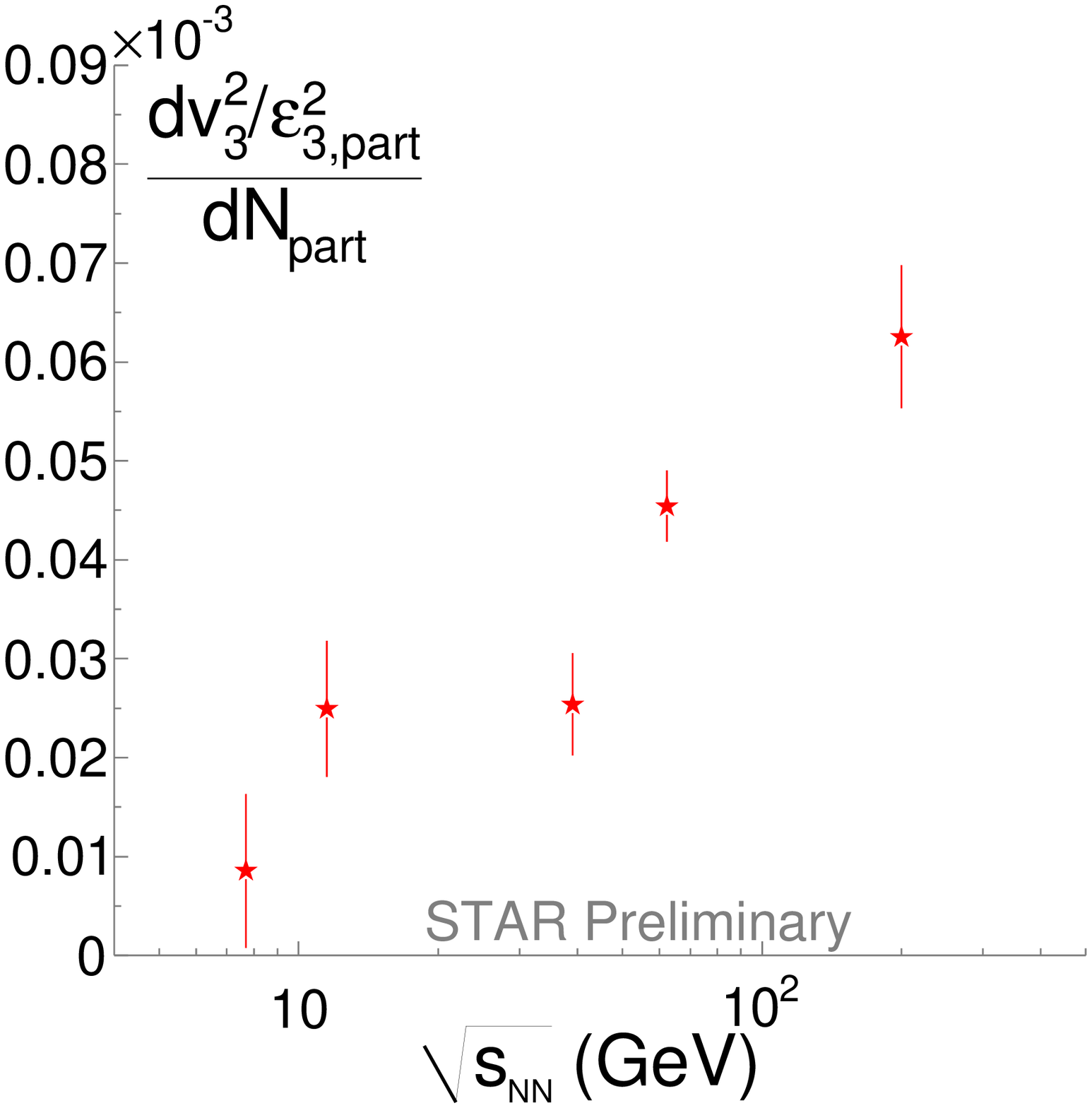}}
\end{center}
\begin{center}
\resizebox{0.40\textwidth}{!}{\includegraphics{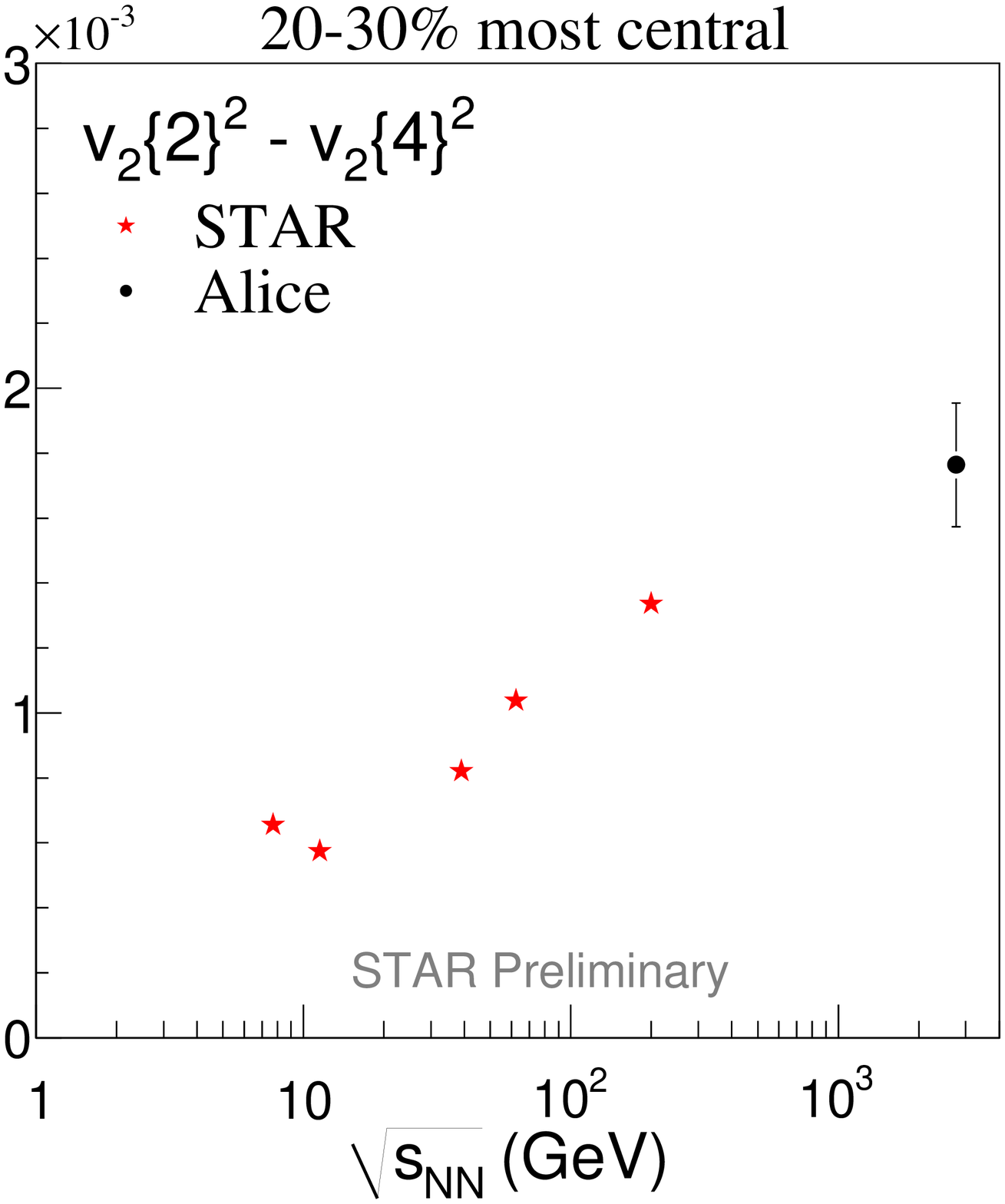}}
\resizebox{0.40\textwidth}{!}{\includegraphics{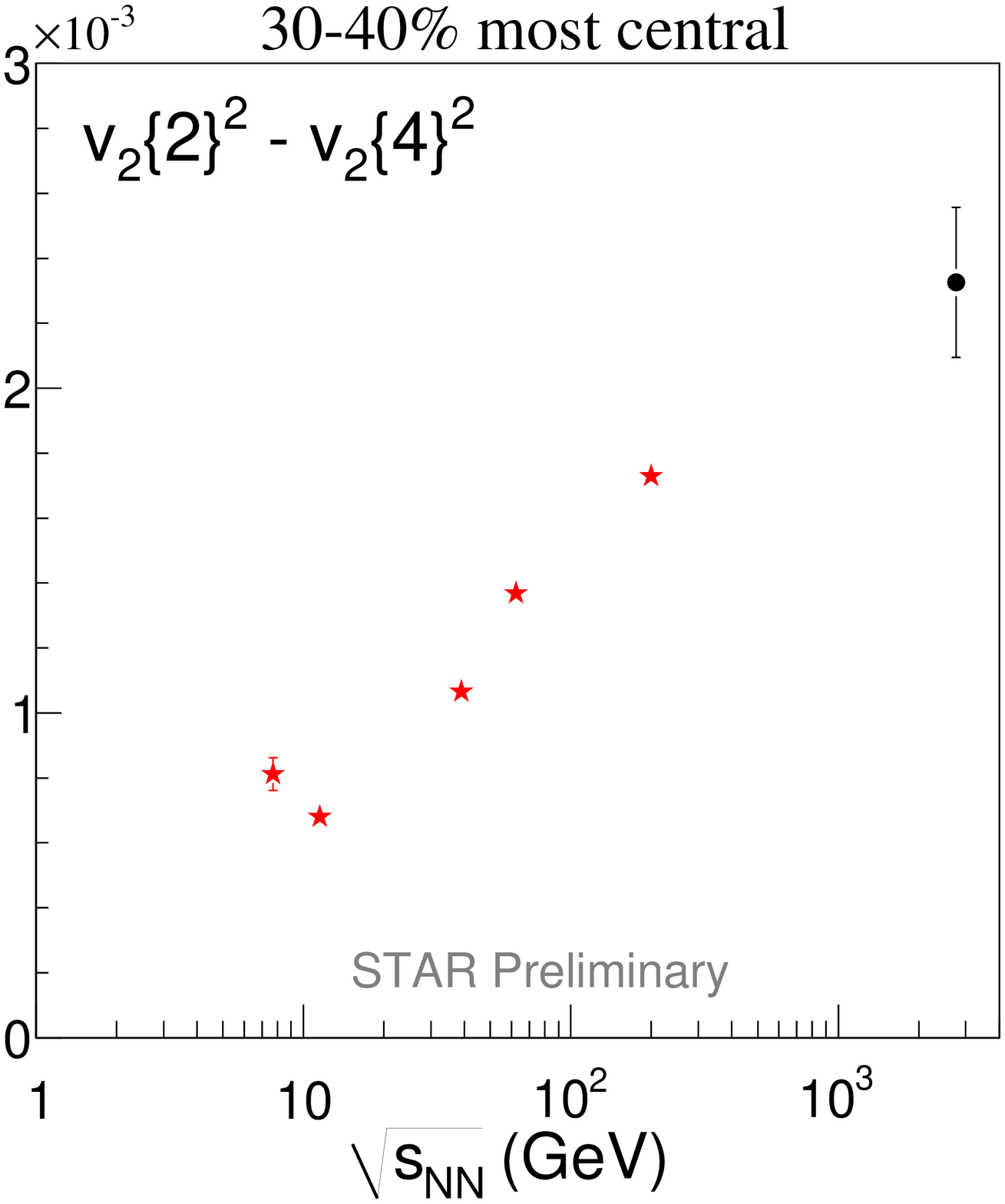}}
\end{center}
\vspace{-0.50cm}
\caption[]{ Top left: $v_3\{2\}^2$ from Q-cumulants with $\eta<1$ scaled by $\varepsilon_{3,part}^2$ vs. centrality at 5 beam energies. Top right: The slope of $v_3\{2\}^2/\varepsilon_{3,part}^2$ vs. $\sqrt{s_{_{NN}}}$. Bottom panels $v_2\{2\}^2-v_2\{4\}^2$ vs. $\sqrt{s_{_{NN}}}$ for two centralities. Bottom panels show statistical errors only (smaller than points). }
\label{f4}
\end{figure}
%\vspace{-0.25cm}

%It is also interesting to study $v_3$ and $v_2$ fluctuations vs. $\sqrt{s_{_{NN}}}$. 
If $v_3$ and $v_2$ fluctuations are created from initial density fluctuations and pressure gradients, one might expect that they are sensitive to the equation-of-state (EOS) of the matter created in the collisions. By scanning $\sqrt{s_{_{NN}}}$, one may be able to vary the initial energy density so that the EOS is soft and the pressure is smaller thus reducing the efficiency of transfer between coordinate and momentum space anisotropies. It remains to be seen, however, if this expectation is bourne out in full models. In Fig.~\ref{f4} we show $v_3\{2\}^2/\varepsilon_{3,part}^2$ vs. $N_{\mathrm{part}}$ for $\sqrt{s_{_{NN}}}=7.7, 11.5, 39, 62.4,$ and 200 GeV. Above $N_{\mathrm{part}}$ of about 50, $v_3\{2\}^2/\varepsilon_{3,part}^2$ rises linearly with $N_{\mathrm{part}}$. The rise increases with $\sqrt{s_{_{NN}}}$. In the top right panel, we show the slope of  $v_3\{2\}^2/\varepsilon_{3,part}^2$ vs. $N_{\mathrm{part}}$ as a function of $\sqrt{s_{_{NN}}}$. The slope shows a generally increasing trend with $\sqrt{s_{_{NN}}}$.  Based on estimates of the Bjorken energy density, the softest point in the EOS should reside near $\sqrt{s_{_{NN}}}=20$ GeV~\cite{eos}. Clearly more energy points will be of use in this study. We have also studied $v_{2}\{2\}^2-v_2\{4\}^2 \approx \delta_2+2\sigma_{v_2}^2$ (see~\cite{mcm}) where $\delta_2$ is the contribution from few-particle correlations unrelated to a common expansion plane (non-flow), and $\sigma_{v_2}$ is the second moment of the $v_2$ distribution which should be sensitive to the initial eccentricity fluctuations~\cite{v2fluc}. $v_{2}\{2\}^2-v_2\{4\}^2$ also increases with $\sqrt{s_{_{NN}}}$ with a possible minimum near $\sqrt{s_{_{NN}}}=20$ GeV. Future studies at 19.6 GeV and 27 GeV~\cite{bes} will help determine whether these measurements are sensitive to the QCD EOS.

\section{Summary}

We presented two and four-particle cumulants of $v_n$ for $n=1,2,3,4,5,$ and 6. The cumulants are consistent with the ansatz $v_n\propto\varepsilon_{n,\mathrm{part}}$. STAR also presented two-particle correlation functions vs. $\Delta\eta$ and $\Delta\phi$ in very central collisions (0-1\%) for particles with $2<p_T<5$~GeV/$c$. Those data exhibit correlation structures consistent with a large contribution from $v_3$ even before background subtraction. Analyses of the $v_n\{2\}$ in central collisions in this $p_T$ range ($p_T>2$~GeV/$c$) indicate that $v_3\{2\}^2>v_2\{2\}^2$ consistent with expectations of models with fluctuations in the initial eccentricity. 
To better understand the higher harmonics, we examined $v_3\{2\}^2$ vs. $\Delta\eta$ and find that it can be fit with a narrow and a wide Gaussian. The narrow Gaussian is consistent with Bose-Einstein correlations and narrows with centrality. The wide Gaussian broadens to a width of $\sigma_{\Delta\eta}=2$. The amplitude of the wide Gaussian first falls with centrality approximately like 1/$N_{\mathrm{part}}$ until $N_{\mathrm{part}}$ of approximately 50, then it scales as $N_{\mathrm{part}}\times\varepsilon_{3,part}^2$. Based on these observations, it appears data favor the ansatz $v_n^2\propto\varepsilon_{n,\mathrm{part}}^2$ with an efficiency for conversion from $\varepsilon_{n,\mathrm{part}}^2$ into $v_n\{2\}^2$ that drops off with $n$ like a Gaussian for low $p_T$ particles but that has a peak at $n=3$ for intermediate $p_T$ particles.% Contributions from non-flow are still under study. %Disentangling correlations from non-flow and from flow is still under investigation.

\end{document}